\documentclass[pra,twocolumn]{revtex4} 
\usepackage{graphicx}
\usepackage{amsmath, amssymb}
\usepackage{color}
\begin{document}

\title{Transport properties for a quantum dot coupled to normal leads 
with pseudogap}

\author{Bunpei Hara}
\affiliation{Department of Physics, Tokyo Institute of Technology, 
Meguro, Tokyo 152-8551, Japan}

\author{Akihisa Koga}
\affiliation{Department of Physics, Tokyo Institute of Technology, 
Meguro, Tokyo 152-8551, Japan}

\author{Tomosuke Aono}
\affiliation{Department of Electrical and Electronic Engineering, 
Ibaraki University, Ibaraki 316-8511, Japan} 
\date{\today}%

\begin{abstract}
We study transport properties for a quantum dot coupled to normal leads 
with a pseudogap density of states at zero temperature,
using the second-order perturbation theory based on the Keldysh formalism.
We clarify that the hybridization function $\Gamma(\omega)\propto |\omega|^r\;
(0 \le r < 1) $ induces the cusp or dip structure in the density of states
in the dot
when finite bias voltage is applied to the interacting quantum dot system.
It is found that the current-voltage characteristics and 
differential conductance are drastically changed at $r=1/2$.
\end{abstract}

\maketitle

Electron transports through the nanoscale systems have attracted
much interest.
One of the fundamental systems is a quantum dot (QD)~\cite{Kouwenhoven},
where the role of electron correlations has been discussed.
Among them, the Kondo effect~\cite{Kondo} 
is one of the central issues in these systems
\cite{GoldhaberGordon,Cronenwett,vanderWiel}
and
it has been clarified that electrons around the Fermi level in leads
screen a localized spin in the QD at low temperatures,
namely the strong coupling (SC) state.
The Kondo screening effect induces
the Kondo resonance peak at the Fermi level,
which yields finite conductance through the QD.
The current-voltage characteristics have been examined 
in terms of 
the perturbation theory with respect to the Coulomb 
energy~\cite{Oguri,Fujii} 
and the the continuous-time quantum Monte Carlo (CTQMC) 
method~\cite{Werner,Werner2}.
The role of the electronic states in the lead in stabilizing the SC state
has been discussed.
When a superconducting lead is attached, the Kondo effect is suppressed 
due to the superconducting gap~\cite{SQN}.
In this case,
the SC state competes with the singlet pairing superconducting states,
which has been discussed experimentally~\cite{Andreev} and 
theoretically~\cite{Fazio,Aono,Domanski,Tanaka,Yamada,KogaQMC}.
It is also interesting to discuss the stability of the SC state 
against the low density of states (DOS) at the Fermi level.

Recently, the systems with a power-law DOS $\rho(\omega) \propto |\omega|^r$
with $r \geq 0$, 
have been realized such as unconventional superconductors,
graphene, and the surface states of the topological insulators.
For instance,
magnetic adatoms on graphene ($r=1$) are responsible for the Kondo effect
~\cite{Wheling,Uchoa}.
This Kondo effect can be controlled by external gate voltages
~\cite{Sengupta,Vojta}, and
the screening effect can be observed using a scanning tunneling microscope.
In addition, a recent experiment shows that the Kondo effect can be induced 
by lattice vacancies~\cite{Chen}.
The magnetic impurity model in fermions with the power-law DOS,
called the pseudogap Kondo model,
shows a competition between the SC and local moment (LM) states
and consequently a quantum phase transition between these states takes place,
which is missing for the conventional Kondo model ($r=0$). 
The pseudogap Kondo problem in equilibrium
has been studied 
intensively~\cite{Withoff,Gonzalez,Bulla,Vojta-Fritz,Bulla2,Glossop3,Fritz-Vojta,Kanao,Aono2},
to understand the nature of the impurity quantum phase transition.

The ground state in equilibrium for the symmetric pseudogap Anderson model
has been discussed.
For $1/2 <r$, the LM state is always realized while
for $0 < r < 1/2$ the SC state is realized
when the Coulomb energy is smaller than
a certain critical value.
The boundary for the impurity quantum phase transition 
between these two states
has been determined by the numerical renormalization group 
method~\cite{Bulla,Gonzalez,Bulla2}.
The perturbative renormalization group (RG) with respect to the $s-d$ coupling 
constant has been applied to determine the RG flow diagram~\cite{Vojta-Fritz}.
The CTQMC method has
been applied 
to discuss finite temperature scaling relations 
in the vicinity of the quantum critical point~\cite{Glossop3}.

It is still nontrivial how the competition 
between the SC and LM states affects the transport properties
at zero temperature although
the universal scaling in electron transport has been found
using the perturbative RG scaling~\cite{Chung}. 
Therefore, it is desired to study how the pseudogap structure affects 
the transport properties in the QD system.
Here we address the issue from the view of the second order perturbation 
(SPT) theory~\cite{Glossop} accompanying
the nonequilibrium CTQMC method.
In the equilibrium system, 
the SC state is not adiabatically connected to the LM state.
Therefore, the simple SPT can not describe the LM state correctly. 
On the other hand, when the bias voltage is applied, 
the DOS at the Fermi energy in both leads is finite, 
which should allow us to perform the SPT theory.
We first check this point by
comparing with the CTQMC results.
Then we discuss in detail the shape of the local density of states
as a function of $r$ to demonstrate a qualitative difference in
the conductance through the QD.

We consider electron transport through a QD 
connected to two leads.
The system is described by the following Anderson impurity Hamiltonian
as
\begin{eqnarray}
\hat{H}&=&
\sum_{k \alpha \sigma}
\left(\varepsilon_k-\mu_{\alpha}\right)
\hat{c}^{\dagger}_{k \alpha \sigma}
\hat{c}_{k \alpha \sigma}+\sum_{k \alpha \sigma}
\left(v_{k\alpha}
\hat{c}^{\dagger}_{k \alpha \sigma}
\hat{d}_{\sigma}
+
\text{h.c.}
\right)\nonumber\\
&+&
\sum_{\sigma}\varepsilon_d
\hat{d}^{\dagger}_{\sigma}
\hat{d}_{\sigma}
+
U\hat{n}_{\uparrow}\hat{n}_{\downarrow},\label{model}
\end{eqnarray}
where $\hat{c}_{k \alpha \sigma} 
(\hat{c}^{\dagger}_{k \alpha \sigma})$ annihilates (creates) an electron
with wave vector $k$ and spin $\sigma(=\uparrow, \downarrow)$ in the 
$\alpha(=L, R)$th lead. 
$\hat{d}_\sigma (\hat{d}^{\dagger}_\sigma)$ annihilates (creates) an electron
at the QD, and 
$\hat{n}_\sigma=\hat{d}^{\dagger}_\sigma\hat{d}_\sigma$.
$\varepsilon_k$ is the dispersion relation of the lead, 
and $v_{k\alpha}$ is the hybridization between the $\alpha$th lead 
and QD. 
$\varepsilon_d$ and $U$ are the energy level and Coulomb interaction 
at the QD, respectively.
To discuss the transport properties in the system, 
we set the chemical potentials as $\mu_{\rm L}=V/2$ and 
$\mu_{\rm R}=-V/2$,
where $V$ is the bias voltage.
For simplicity, we focus on the particle-hole symmetric case with 
$\varepsilon_d+U/2=0$.
We consider the following hybridization function,
\begin{align}
\Gamma(\omega)=\pi\sum_{k\alpha} |v_{k\alpha}|^2\delta(\omega-\varepsilon_{k})
=\Gamma\left|\frac{\omega}{\Gamma}\right|^r,
\end{align}
with a constant $\Gamma$ and $ 0 \leq r < 1$. 
Note that when $V\neq0$, $\Gamma(\mu_{\rm L}) = \Gamma(\mu_{\rm R}) \neq 0$
while $\Gamma(\mu_{\rm L}) = \Gamma(\mu_{\rm R}) = 0$ at zero voltage.
All energy scales are measured in units of $\Gamma$ below.

In the paper, we apply the SPT to the model Hamiltonian (\ref{model})
using the Keldysh Green function method.
The selfenergy is then given as~\cite{Hershfield,Oguri},
\begin{align}
\Sigma^\alpha_{\sigma}
=-U^2
\int\!\!\!\!\int
\frac{d\omega_{1}}{2\pi}
\frac{d\omega_{2}}{2\pi}
{\cal G}^{\alpha}_{\bar{\sigma}}(\omega_{1})
{\cal G}^{\bar{\alpha}}_{\bar{\sigma}}(\omega_{1}+\omega_{2}-\omega)
{\cal G}^{\alpha}_{\sigma}(\omega_{2}),
\end{align}
where ${\cal G}_{\sigma}^{\alpha}\;(\alpha=<,>)$ is 
the noninteracting Keldysh Green's function. 
To study the electron transport through the QD,
we first calculate the DOS in the QD, which is given as
\begin{align}\label{eq;DOS}
\rho(\omega)&=-\frac{1}{\pi}\;{\rm Im}
\frac{1}{\omega-\Delta^r(\omega)-\Sigma^r(\omega)},
\end{align}
where $\Delta^r=-\left[\tan (\pi r/2) \;{\rm sgn}\;\omega+i\right]\Gamma(\omega)$~\cite{Gonzalez,Glossop}
and $\Sigma^r$ is the retarded selfenergy.
Its real and imaginary parts are given 
as
\begin{align}
&{\rm Re}\;\Sigma^{r}_\sigma(\omega)=\frac{1}{\pi} {\cal P} 
\int \frac{{\rm Im}\; \Sigma^r_\sigma(x)}{x-\omega}dx,\\
&{\rm Im}\;\Sigma^{r}_\sigma(\omega)=
\frac{i}{2}\Big[\Sigma^>_\sigma(\omega)-\Sigma^<_\sigma(\omega)\Big],
\end{align}
where ${\cal P}$ is the Cauchy's principal-value integral.
Calculating the current $I=\int_0^{V/2} \rho (\omega) \Gamma(\omega) d\omega$
and the differential conductance $G(V)=\partial I / \partial V$ numerically,
we discuss how
$\rho(\omega)$ and $G(V)$ depend on $U$ and $r$.

\begin{figure}[htb]
\begin{center}
\includegraphics[width=7.5cm]{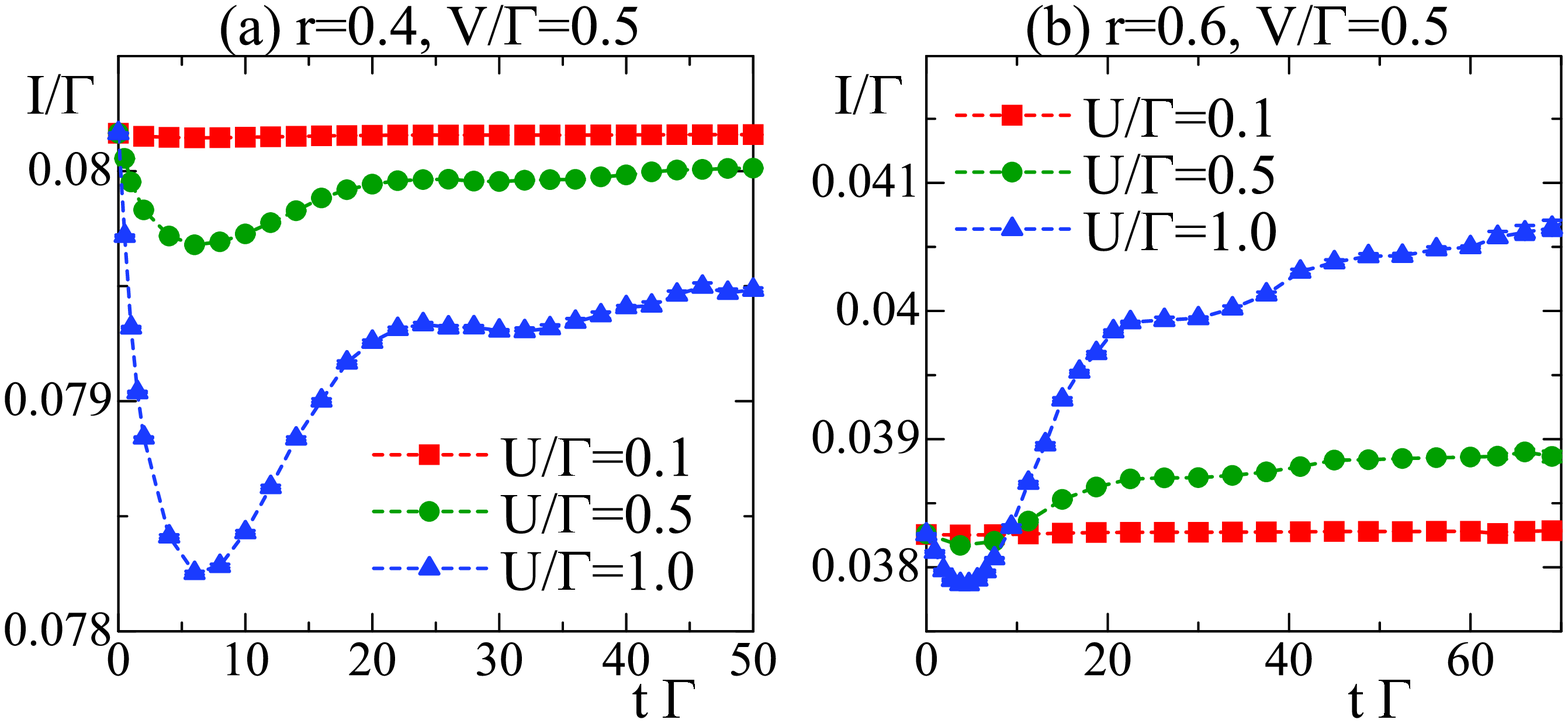}
\caption{(Color online)
The time evolutions of the current $I$ at $V/\Gamma=0.5$ and $T=0$
when $r=0.4$ (a) and $r=0.6$ (b). 
The solid squares, circles, and triangles 
represent $I$ with 
$U/\Gamma=0.1, 0.5$, and $1.0$, respectively.
}
\label{fig:qmc}
\end{center}
\end{figure}

\begin{figure}[htb]
\begin{center}
\includegraphics[width=7.5cm]{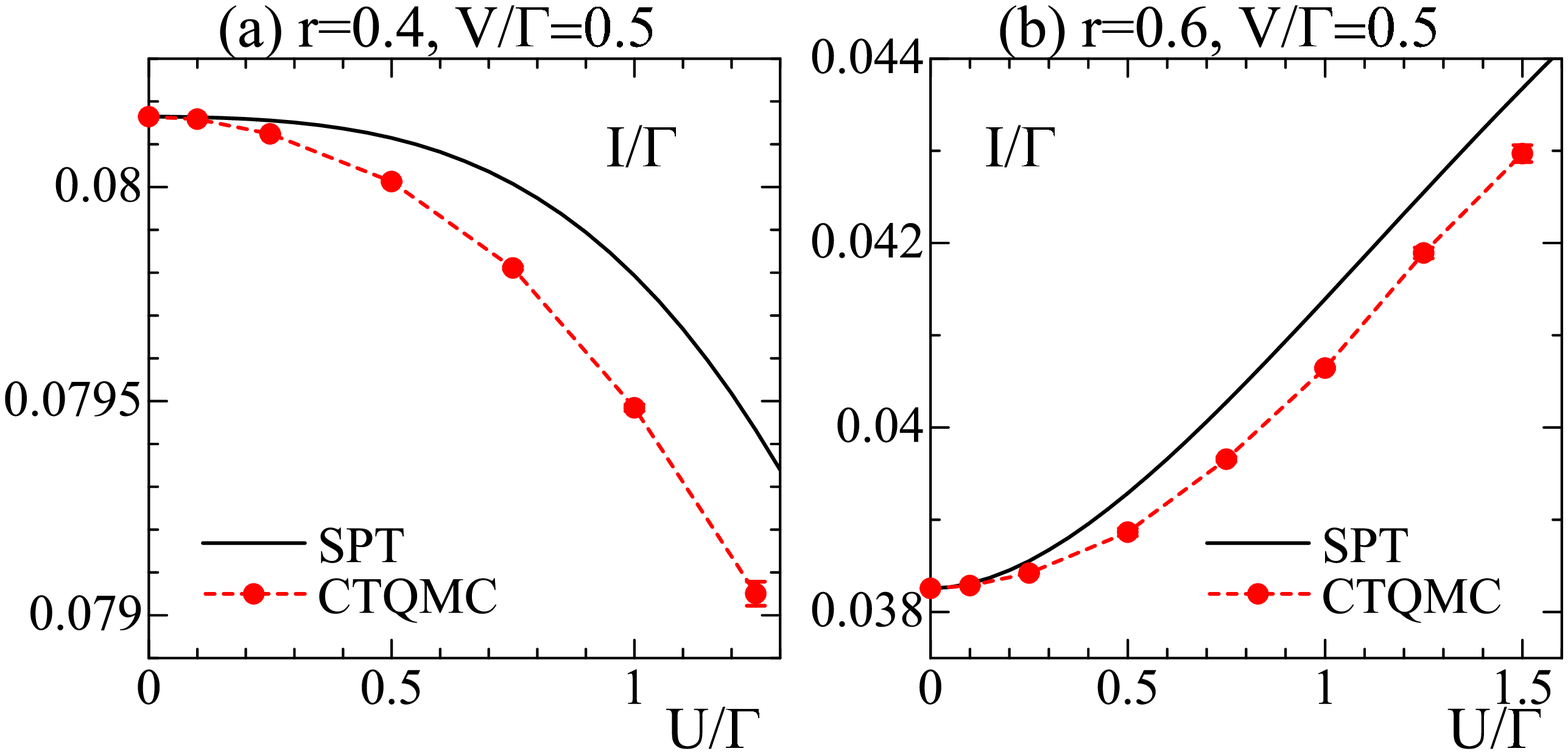}
\caption{(Color online)
The current $I$ as a function of $U$ 
at $V/\Gamma=0.5$ and $T=0$
when $r=0.4$ (a) and $r=0.6$ (b). 
The solid lines represent the results obtained from the SPT.
The open circles are obtained by the nonequilibrium CTQMC method 
with a fixed maximum time $t_{\rm max}\Gamma=50.0$ ($69.0$) 
for the system with $r=0.4 (0.6)$.
}
\label{fig:qmc+2nd}
\end{center}
\end{figure}

Before we proceed our discussion, we check the validity 
of the SPT.
To this end, we also calculate $I$ using the CTQMC method
in the continuous-time auxiliary field formation~\cite{Werner,Werner2}.
In the method, one can examine the time evolution of physical quantities
after the interaction quench.
Figure~\ref{fig:qmc} shows the time evolution of $I$
in the systems with $r=0.4$ and $0.6$.
When $U/\Gamma \leq 0.5$, 
$I$ is slightly changed from the initial value and converges 
to a certain value.
Then we can regard that the system reaches the steady state. 
On the other hand, when $U/\Gamma=1.0$, $I$ gradually increases,
as shown in Fig. \ref{fig:qmc}.
This originates from the fact that the low DOS around $\omega=0$ 
increases the characteristic time scale of the screening effect.
This results in a slow decay to the steady state.
Therefore, it is hard to deduce the steady current precisely
within the restricted numerical resources,
in particular, in the large $U$ regime.
However, the CTQMC method allows us to study the tendency of 
the physical quantities
correctly if a certain time $t_{\rm max}$ is appropriately chosen
beyond the initial relaxation time.
The obtained results with a fixed time $t=t_{\rm max}$
are shown in Fig.~\ref{fig:qmc+2nd}.
When $r=0.4 (0.6)$, the increase of $U$ decreases (increases) $I$.
The CTQMC results are in good agreement with the SPT ones;
even when $U/\Gamma=1.5$, the discrepancy between the two methods 
is within a few percents.
Note that the consistency for $r=0.6$ is in contrast to that for 
the zero bias voltage case
since it is known that
the SPT method is not applicable 
when $1/2 < r < 1$ for zero bias voltage~\cite{Glossop}.
These results indicate that the SPT correctly describes
the QD system with the power-law hybridization $(0 \leq r < 1)$ at finite bias voltage.

\begin{figure}[htb]
\begin{center}
\includegraphics[width=6.5cm]{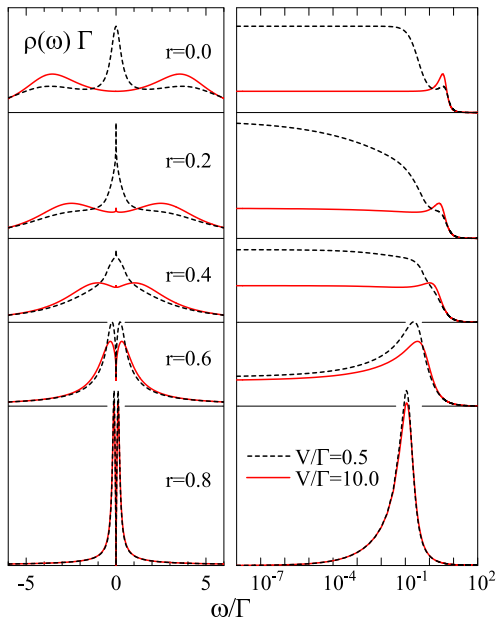}
\caption{(Color online)
The DOS plots for the QD are shown in linear (left panel) and 
logarithmic (right panel) scales 
in the system with $U/\Gamma=8.0$ and $r=0.0, 0.2, 0.4, 0.6$, and $0.8$, 
respectively. 
}
\label{fig:dos}
\end{center}
\end{figure}

First, we study the shape of $\rho(\omega)$ as a function of $r$ using the SPT.
Figure~\ref{fig:dos} shows the results for $U/\Gamma=8.0$.
When $r=0$ and $V/\Gamma=0.5$, 
the Kondo resonance peak
appears at $\omega\sim 0$ together with the Coulomb peaks 
at $\omega\sim \pm U/2$.
As the bias voltage increases beyond the Kondo energy scale,
the Kondo resonance peak smears out and two Coulomb peaks 
are visible~\cite{Fujii,Taniguchi}.
As $r$ increases, these two peaks approach each other.
For instance,
the peaks are located at $\omega\sim \pm 0.5\Gamma$
when $r=0.6$.
Since its energy is far from the bare Coulomb interaction, 
one may regard them as the {\it renormalized} Coulomb peaks.
In addition to them, 
we find the cusp or dip structure around $\omega=0$ induced by 
the introduction of $r$.
This behavior is clearly found in the logarithmic plot of $\rho(\omega)$,
as shown in the right panel of Fig.~\ref{fig:dos}.
In the case $0<r<0.5$ with the low bias voltage $V/\Gamma=0.5$,
the sharp cusp structure appears together with the Kondo resonance peak.
This is similar to the low-energy structure 
in the equilibrium system ($V=0$) with $r>0$~\cite{Bulla},
which suggests that the steady state with $0<r<0.5$ and $V/\Gamma=0.5$
should be characterized by the SC state.
We also find that 
the sharp peak structure remains even in the large $V$ case.
Therefore, this sharp peak is distinct from the Kondo resonance one.
When $1/2 < r < 1$, on the other hand,
the dip structure appears, namely the absence of the Kondo resonance.

To clarify how the cusp or dip structure appears in $\rho(\omega)$, 
we numerically examine the selfenergy around $\omega=0$
and then find the following power-law dependence:
\begin{align}\label{eq:selfenergypower}
\begin{array}{rcl}
{\rm Re}\Sigma^{r}(\omega) &\sim&
\left\{
\begin{array}{ll}
\displaystyle a\frac{\omega}{\Gamma}, & 0\leq r<1/3, \\
\displaystyle a\;\text{sgn}(\omega)\left|\frac{\omega}{\Gamma}\right|^{2-3r},  & r>1/3,
\end{array}
\right. \\ 
{\rm Im}\Sigma^{r}(\omega) &\sim&
\displaystyle -b+c\left|\frac{\omega}{\Gamma}\right|^{2-3r},
\end{array}
\end{align}
where the real coefficients $a, b$ and $c$
are certain constants.
Substituting Eq.~(\ref{eq:selfenergypower}) into Eq.~(\ref{eq;DOS}),
the power-law behavior of $\rho(\omega)$ around $\omega=0$ is classified into five cases:
\begin{equation}
\rho(\omega) \sim
\left\{
\begin{array}{ll}
\displaystyle
\frac{1}{\pi(\Gamma+b)}
\left[
1+O\left(\left(\frac{\omega}{\Gamma}\right)^2\right)
\right],
& r=0,   \\ 
\displaystyle
\frac{1}{\pi b}
\left[
1-
\frac{\Gamma}{b}
\left|\frac{\omega}{\Gamma}\right|^{r}
\right],
& 0<r<1/2, \label{asy1}  \\
\displaystyle
\frac{1}{\pi b}
\left[
1+
\frac{c-\Gamma}{b}
\left|\frac{\omega}{\Gamma}\right|^{\frac{1}{2}}
\right],
& r=1/2, \\
\displaystyle
\frac{1}{\pi b}
\left[
1+
\frac{c}{b}
\left|\frac{\omega}{\Gamma}\right|^{2-3r}
\right],
& 1/2<r<2/3, \\
\displaystyle
\frac{1}{\pi c}
\cos^2\left[\frac{\pi}{2}(2-3r)\right]
\left|\frac{\omega}{\Gamma}\right|^{3r-2},
& 2/3<r.
\end{array}
\right.
\end{equation}
When $r=0$, the system is reduced to the conventional QD, 
where $\rho(\omega)-\rho(0)\sim O(\omega^2)$.
When $U>0$, $\rho(0)$ is finite for $0< r <2/3$ while
$\rho(0)=0$ for $2/3 < r < 1$.
When $0<r<1/2$, 
$\rho(\omega)$ always has a local maximum at $\omega=0$
since $\Gamma/b^2  > 0$.
On the other hand, in the case $1/2<r$, we find $c>0$,
and a dip structure appears around $\omega=0$.
When $r=1/2$, the low energy structure of $\rho(\omega)$ depends on $U$
since it determines the ratio $c/\Gamma$. 
\begin{figure}[htb]
\begin{center}
\includegraphics[width=5.5cm]{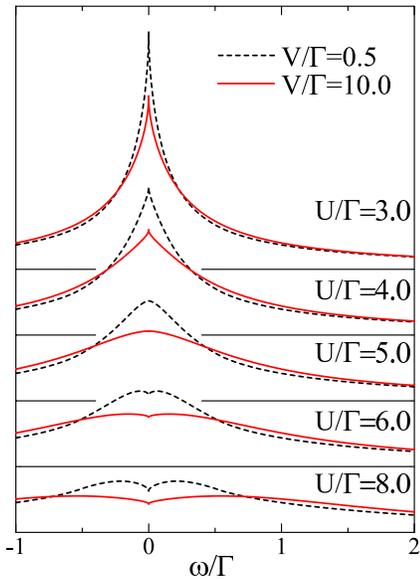}
\caption{(Color online)
The DOS plots for the system with $r=1/2$.
}
\label{fig:dos05}
\end{center}
\end{figure}
Figure~\ref{fig:dos05} shows $\rho(\omega)$ with $r=1/2$
for various values of $U$.
When $U<U_0\; (U>U_0)$, the cusp (dip) structure appears,
where $U_0$ is the critical interaction at which $c/\Gamma=1$.
We find $U_0/\Gamma\sim 5$
with little dependence on $V$.
\begin{figure}[htb]
\begin{center}
\includegraphics[width=6.5cm]{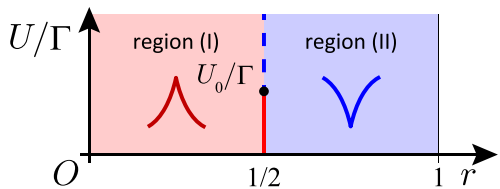}
\caption{(Color online)
Phase diagram
when the finite bias voltage is applied.
In the region I (II), the cusp (dip) structure appears in $\rho(\omega)$
around $\omega=0$.
}
\label{fig:diagram}
\end{center}
\end{figure}

We end up with the schematic diagram for $\rho(\omega)$ around $\omega=0$, 
as shown in Fig.~\ref{fig:diagram}.
We wish to note that, in the case (I) with $0<r<1/2$, 
the prefactor of the power-law $\Gamma/b^2$ indicates that
the cusp structure is formed by 
the cooperative phenomenon between the interaction and the hybridization.
Therefore, it is expected that the steady state denoted by the region (I) 
should be continuously connected to the SC state, 
which is realized in the equilibrium system with $r<1/2$ and small $U$ case.
On the other hand, in the case (II) with $r>1/2$, 
the prefactor
depends only on the selfenergy.
Therefore, we can say that the steady state denoted by the region (II) is
connected to the LM state with the gap when $V\rightarrow 0$.
Then we find that the obtained diagram is consistent with
the ground state phase diagram~\cite{Bulla,Gonzalez,Bulla2}.
This means that the SPT based on 
the Keldysh formalism captures the essence of ground state properties.

\begin{figure}[htb]
\begin{center}
\includegraphics[width=6.5cm]{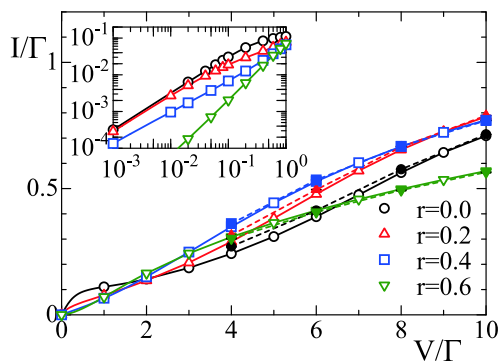}
\caption{(Color online)
The current $I$ as a function of the bias voltage $V$
for $U/\Gamma=8.0$.
Open symbols represent the SPT results.
Solid symbols represent the results
by the CTQMC method.
}
\label{fig:I}
\end{center}
\end{figure}

Next, we discuss the transport properties in the system.
In Fig.~\ref{fig:I}, we show the current $I$ obtained by the SPT and CTQMC
methods for $U/\Gamma=8.0$.
In the high voltage regime, it displays the excellent agreement 
between the two methods.
In the low voltage regime, it is hard to obtain the steady current
by means of the CTQMC method due to the serious sign problem.
The SPT results exhibit that the current strongly depends on the power $r$,
as shown in the inset of Fig. \ref{fig:I}.
Namely, the introduction of the bias voltage linearly increases 
the current in the case $r<1/2$, 
while $I$ is suppressed for $r > 1/2$.

\begin{figure}[htb]
\begin{center}
\includegraphics[width=6.5cm]{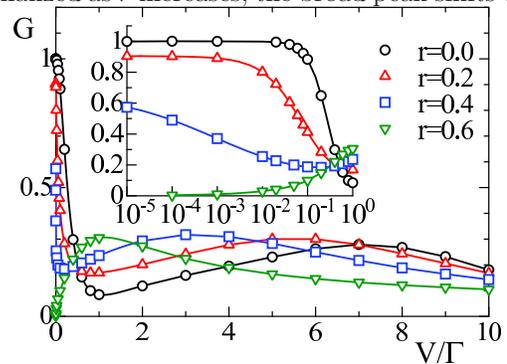}
\caption{(Color online)
The differential conductance $G$ as a function of the bias voltage $V$
in the system with $U/\Gamma=8.0$.
}
\label{fig:G}
\end{center}
\end{figure}

To make these clear,
we also calculate the differential conductance $G$, as shown in Fig.~\ref{fig:G}.
At large voltages,
$G$ show a broad peak determined by 
the high energy Coulomb peaks in $\rho(\omega)$.
Since the Coulomb energy is renormalized as $r$ increases,
the broad peak shifts toward lower $V$.
On the other hand, $G$ around $V=0$
is affected by the low energy structure in $\rho(\omega)$.
When $r=0$, the Kondo peak yields the unitary limit of $G$
at $V=0$.
Increasing the power $r$, the Kondo peak becomes obscure and
the cusp behavior is induced instead, as discussed above.
The sharp cusp leads to the sudden decrease on the introduction of 
the voltage, as shown in Fig.~\ref{fig:G}.
On the other hand, when $r>0.5$, the dip structure in $\rho(\omega)$
suppress $G$ around $V$.
This is in contrast to the result for $r < 1/2$
although $G$ at $V=0$ can not be estimated directly.
To see this difference clearly,
we show the logarithmic plot of $G$
in the inset of Fig.~\ref{fig:G}.
It is found that, in the case $r\le 0.4$, 
$G$ is finite
while it vanishes in the case $r=0.6$.
In this way,
the $G$-$V$ characteristics can display
the difference between the regions (I) and (II)
in Fig.~\ref{fig:diagram}.
It is an interesting problem how stable $r$-dependence transport properties
are at finite temperatures, which is now under consideration.

We have investigated transport properties through a QD coupled 
to normal leads with the pseudogap density of states.
Using the second-order perturbation theory based on the Keldysh formalism, 
we have examined the DOS for the QD.
The phase diagram with the LM and SC states is shown for finite bias voltage.
The $G$-$V$ characteristics distinguish between LM and SC states 
in the phase diagram.

\begin{acknowledgements}
This work was partly supported by the Grant-in-Aid 
for Scientific Research from JSPS, KAKENHI No. 25800193. (A.K.)
Some of computations in this work have been
carried out, by using the TSUBAME2.5 supercomputer 
in the Tokyo Institute of Technology
through the HPCI System Research Project (Project ID:hp140153).
The simulations have been performed, 
by using some of ALPS libraries~\cite{alps1.3}.
\end{acknowledgements}

\end{document}